# Intracellular micro-rheology probed by micron-sized wires

L. Chevry, R. Colin, B. Abou and J.-F. Berret*

*Laboratoire Matière et Systèmes Complexes (MSC), UMR 7057 CNRS &Université Paris Diderot, Bâtiment Condorcet, 10 rue Alice Domon et Léonie Duquet, 75205 Paris (France)*

**Abstract : In the last decade, rapid advances have been made in the field of micro-rheology of cells and tissues. Given the complexity of living systems, there is a need for the development of new types of nano- and micron-sized probes, and in particular of probes with controlled interactions with the surrounding medium. In the present paper, we evaluate the use of micron-sized wires as potential probes of the mechanical properties of cells. The wire-based micro-rheology technique is applied to living cells such as murine fibroblasts and canine kidney epithelial cells. The mean-squared angular displacement of wires associated to their rotational dynamics is obtained as a function of the time using optical microscopy and image processing. It reveals a Brownian-like diffusive regime of the form $\langle \Delta \psi^2(t,L) \rangle \sim t/L^3$, where $L$ denotes the wire length. This scaling suggests that an effective viscosity of the intracellular medium can be determined, and that in the range $1 - 10$ μm it does not depend on the length scale over which it is measured.**

## 1. Introduction

Viscosity is one of the major parameters determining the diffusion rate of species in condensed media, and in particular in cells and tissues. In living organisms, studies have shown that viscosity changes are linked to disease and dysfunction at the cellular level or to developmental malformation at the level of tissues [1, 2]. These perturbations are caused by changes in mobility of chemicals within the cell, affecting fundamental processes such as membrane trafficking, formation of protein scaffolds and maintenance of cell shape. Macromolecular crowding has been shown to have a significant impact on the rates and the equilibria of interactions between biological molecules in intracellular environments [3]. In the last decade, cell mechanics has attracted much interest, and rapid advances have been made in the fields of micro-rheology of cells and that of tissues [1, 4-7]. Among the techniques developed to probe the active and passive responses of the cytoskeleton, two classes have been presented: the first involves techniques such as atomic force microscopy, laser and magnetic tweezers and pipette suction which explore the mechanical response via the cell surface [8-11]. The second class relies on the tracking of probe particles, organelles or molecules by optical microscopy [12-19]. For the tracking of embedded probes, a majority of experiments were performed with $0.1 - 1$ μm beads internalized *via* endocytosis.

Authors have found that the mechanical response of cells obey scaling laws for time or frequency dependencies. The mean-squared displacement (MSD) of internalized probes, such as micro- or submicrometer beads scales as $\langle \Delta r^2(t) \rangle \propto t^a$, where $a$ is an exponent that depends on the time scale [4, 20]. At short times (t < 1s), subdiffusive motions characterized by scaling exponents below 1 were found, whereas at longer times superdiffusive dynamics coincide with intracellular active transport or macroscopic motions of the cells [4, 5, 13, 14, 16, 19]. For many authors, the scaling found for the MSD or for the moduli are the signature that cells behave as a soft glassy material [10, 12]. An additional feature of cell mechanics is that their interior presents strong structural heterogeneities, which affect the cell-to-cell and tracer-to-tracer reproducibility, and produce dramatic variations of the measured mechanical quantities [4, 5, 13-17, 21, 22].

In view of the contrasting results obtained with spherical beads in the evaluation of cell mechanics [4, 7, 15], and also considering the complexity of living system dynamics, there is a need to develop new types of nano- and micron-sized probes. In a recent study, fluorescent molecules described as molecular rotors were designed to exhibit a non-radiative decay of emitted light that depends on the viscosity of the surrounding medium [2]. Applications to living cell interiors (SK-OV-3 ovarian carcinoma cells) provided a mapping of the intracellular viscosity and an average viscosity of 0.14 Pa. s *i.e.* 170 times the viscosity of water at body temperature. Several studies have also emphasized the possibility of exploiting anisotropic objects such as disks [23], rods [17, 24-28] and wires [29-33] for both passive [23, 32, 33] and active [17, 25-27, 29-31] microrheology. As compared to spherical beads, anisotropic probes were hardly investigated, one of the reasons being the difficulty to quantify their three-dimensional (3D) motions by microscopy [24]. Recently, Xiao *et al.* have demonstrated that the translational and rotational dynamics of 67 nm long gold nanorods could be resolved by coupled planar illumination microscopy imaging, a method that give access to the particle trajectories and orientations [28].

In the present study, a wire-based micro-rheology technique was employed to probe the mechanical response of the intracellular medium of living murine fibroblasts and canine kidney epithelial cells. Highly persistent wires of diameter 400 nm and length between 1 μm and 10 μm were synthesized by co-assembly [34, 35] and internalized into living cells. The present work was motivated by the benefits shown by micron-sized wires as probes of the intracellular medium. The wires were easily uptaken by cells such as 2139 human lymphoblasts, NIH/3T3 murine fibroblasts and canine kidney epithelial cells [36]. Interestingly, the wires were found to be disperse in the cytosol, a situation that enables them to probe directly the cytoskeleton and filament network [5, 26, 36]. As a proof of concept, the technique described here was conducted on purely viscous fluids and showed excellent results for the determination of the static viscosity [33].

# 2. Materials and Methods

## 2.1. Magnetic wires synthesis

Wires were formed by electrostatic complexation between oppositely charged nanoparticles and copolymers [34, 35]. The particles were 8.3 nm iron oxide nanocrystals (γ-$Fe_2O_3$, maghemite) synthesized by polycondensation of metallic salts in alkaline aqueous media [37]. Extensive characterization of the particles including their size distributions, surface charges, structural anisotropy and magnetization can be found in Supporting Information (S1 – S3). To improve their colloidal stability, the cationic particles were coated with $M_W = 2100 \ g \ mol^{-1}$ poly(sodium acrylate) (Aldrich) using the precipitation-redispersion process [38]. This process resulted in the adsorption of a highly resilient 3 nm polymer layer surrounding the particles. The copolymer used for the wire synthesis (S4) was poly(trimethylammoniumethylacrylate)-*b*-poly(acrylamide) with molecular weights 11000 $g\ mol^{-1}$ for the charged block and 30000 $g\ mol^{-1}$ for the neutral block [39, 40]. Fig. S4 displays transmission optical and electron microscopy images of the γ-Fe$_2$O$_3$ wires. As shown in the figure, the wires are polydisperse. Their length distribution was described by a log-normal function with median length $L_0$ and polydispersity $s_{NW}$. Throughout the manuscript, the polydispersity is defined as the ratio between the standard deviation and the average value. The shelf life of the co-assembled structures is of the order of several years. For passive micro-rheology experiments conducted on model fluids (water/glycerol mixtures), the wire sample was characterized by a median length $L_0 = 27\ \mu m$ and a polydispersity $s_{NW} = 0.65$. For this sample, the lengths were comprised between 1 and 100 μm. For the experiments on fibroblasts, wires were synthesized at $L_0 = 4\ \mu m$ and $s_{NW} = 0.50$ [36]. Studied by scanning electron microscopy, the average diameter $d$ of the wires was estimated at 400 nm [33]. Electrophoretic mobility and ζ–potential measurements made with a Zeta sizer Nano ZS Malvern Instrument showed that the wires were electrically neutral [35].

## 2.2. Cell culture

NIH/3T3 fibroblast cells from mice were grown in T25-flasks as a monolayer in DMEM with high glucose (4.5 g L$^{-1}$) and stable glutamine (PAA Laboratories GmbH, Austria). This medium was supplemented with 10% fetal bovine serum (FBS) and 1% penicillin/streptomycin (PAA Laboratories GmbH, Austria), referred to as cell culture medium. Exponentially growing cultures were maintained in a humidified atmosphere of 5% CO$_2$ and 95% air at 37°C, and in these conditions the plating efficiency was 70 – 90% and the cell duplication time was 12 – 14 h. Cell cultures were passaged twice weekly using trypsin–EDTA (PAA Laboratories GmbH, Austria) to detach the cells from their culture flasks and wells. The cells were pelleted by centrifugation at 1200 rpm for 5 min. Supernatants were removed and cell pellets were re-suspended in assay medium and counted using a Malassez counting chamber. Cellular growth was measured with both untreated cells and cells treated for 24 h with different concentrations of 4 μm nanowires ranging from 3 to 60 wires per cell.

The MDCK cell line was derived from the kidney of an adult female cocker-spaniel by S. H. Madin and N. B. Darby. The culture medium was Dulbecco's modified Eagle's medium (DMEM) with 10% heat-inactivated fetal bovine serum and antibiotics. Cells seeded at a concentration of $2 \times 10^4$ cells cm$^{-2}$ in the above culture medium were 100% confluent in 5 days. After a phase of 1-2 days during which the cell concentration remained stable, MDCK entered the exponential growth phase with a mean doubling time of 12 hours. Although the cells adhere very strongly to each other, as well as to the plastic substrates, it was possible to passage them using a trypsin for 15 mn. The MDCK cells were treated with iron oxide based wires using the same protocol as for the NIH/3T3. Cells were incubated 24 h at a concentration of 10 wires per cell. The observation of their orientation fluctuations in the cytoplasm was made by phase-contrast or bright field microscopy. The data analysis and treatment were similar to those used for fibroblasts.

## 2.3. Transmission optical microscopy

Phase-contrast images of the cells containing wires were acquired on an IX71 inverted microscope (Olympus) equipped with 40× and 60× objectives. $2 \times 10^5$ NIH/3T3 fibroblast cells were first seeded onto 3.6 cm Petri dishes for 24 h prior incubation with wires. μl-aliquots containing wires were then added to the supernatant. The wire concentration was determined by the ratio of the number of wires incubated per cell, here fixed at 10. The incubation of the wires lasted 24 hours. On the third day, excess medium was removed and the cells were washed with a PBS solution (with calcium and magnesium, Dulbecco's, PAA Laboratories), trypsinized and centrifuged. Cell pellets were re-suspended in Dulbecco's modified Eagle's medium (DMEM) complemented with HEPES at 1.5%. For optical microscopy, 20 μl of the previous cell suspension were deposited on a glass plate and sealed into to a Gene Frame® (Abgene/Advanced Biotech) dual adhesive system. The sample was then left for 4 h in the incubator to let cells adhere onto the glass plate. The image acquisition system consisted of a Photometrics Cascade camera (Roper Scientific) working with Metaview (Universal Imaging Inc.). In order to determine the length distribution of the wires, pictures were digitized and treated by the ImageJ software (http://rsbweb.nih.gov/ij/).

## 2.4. Transmission Electron Microscopy

NIH/3T3 fibroblast cells were seeded onto the 6-well plate. After a 24 h incubation with 4 µm wires, the excess medium was removed. The cells were washed in 0.2 M phosphate buffer (PBS) and fixed in 2% glutaraldehyde-phosphate buffer (0.1 M) for 1 h at room temperature. Fixed cells were further washed in 0.2 M PBS. The cells were then postfixed in a 1% osmium-phosphate buffer for 45 min at room temperature in dark conditions. After several washes with 0.2 M PBS, the samples were dehydrated by addition of ethanol. Samples were infiltrated in 1:1 ethanol:epon resin for 1 h and finally in 100% epon resin for 48 h at 60°C for polymerization. 90 nm-thick sections were cut with an ultramicrotome (LEICA, Ultracut UCT) and picked up on copper-rhodium grids. They were then stained for 7 min in a 2%-uranyl acetate and for 7 min in a 0.2%-lead citrate. Grids were analyzed with a transmission electron microscope (ZEISS, EM 912 OMEGA) equipped with a $LaB_6$ filament, at 80 kV. Images were recorded with a digital camera (SS-CCD, Proscan 1024×1024), and the iTEM software.

## 2.5. Optical microrheology

An inverted Leica DM IRB microscope with a ×100 oil immersion objective (NA = 1.3), coupled to a fast digital camera (EoSens Mikrotron) was used to record the 2D projection of the wires on the focal plane objective. To prevent hydrodynamic coupling, the wire concentration was chosen around c = 0.01 wt. % and the tracking was made far enough from the walls of the observation chamber. The microscope objective temperature was controlled within 0.1 °C using a Bioptechs heating ring coupled to a homemade cooling device. The sample temperature was controlled through the immersion oil. Sedimentation of the wires was negligible on the recording time scales. The camera was typically recording 100 images per second over 100 s (corresponding to 10000 images). The 3D Brownian motion of the wires was extracted from their 2D projection on the $(x, y)$-plane, according to a procedure described previously [33] (S5). The angle $\varphi(t)$ and the apparent length $L_{app}(t)$ were measured from the images, using a homemade tracking algorithm implemented as an ImageJ plugin. The MSAD, reflecting the out-of-plane Brownian motion of the wires, was then computed from $\varphi(t)$ and $L_{app}(t)$.

# 3. Results and discussion

## 3.1. Internalization of the wires in the NIH/3T3 murine Fibroblasts

The interactions between the wires and NIH/3T3 mouse fibroblasts were investigated to explore the possibilities of using the wires for micro-rheology and cell manipulation [36]. The major finding of this previous research was that after an incubation of 24 h, a large proportion of wires crossed the plasma membrane spontaneously without compromising the cell viability and cell growth. The results on the internalization and localization of the wires in living cells are as follows.

Wires dispersed into the cytoplasm were directly observed by optical microscopy. As shown in Fig. 1a, elongated threads of various lengths and orientations are visible in the perinuclear region of a single cell. Electron microscopy of treated cells confirmed that the wires were located either in membrane-bound compartments or dispersed in the cytosol, the latter occurring more frequently. Examples of elongated nanostructures not limited by a membrane are highlighted in Fig. 1b. Wires were not observed inside the nucleus [26] or stuck at the plasma membrane. When exposed to a polydisperse population of wires, it was found that the cells sorted the wires according to their length. Studies show that the internalization was effective only for lengths of the order or lower than the average size of the adherent fibroblasts, *i.e.* $10 - 15$ µm [41].

optical microscopy

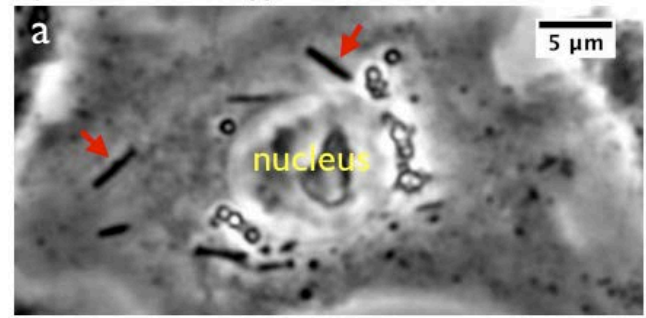

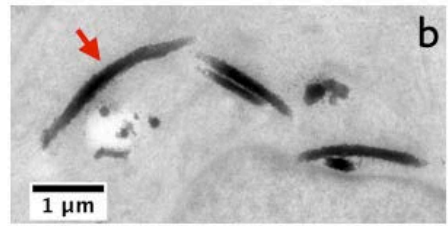
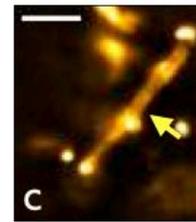

*Figure 1: a)* *Phase contrast microscopy images of NIH/3T3 fibroblasts cells treated with 4 μm wires for 24 h at a concentration of 10 wires per cell. Wires inside the cytoplasm are indicated by red arrows.* *b)* *Representative transmission electron microscopy image of NIH/3T3 fibroblasts incubated in the same conditions. The wires also appeared to be shorter than their initial length and exhibited sharp and diffuse extremities. These observations were attributed to the fact that the microtomed sections of cells were 90 nm thick, and that wires outside the plane of the cut were shortened by the sample preparation.* *c)* *Cells prepared in the same conditions as in (a) and (b) were fixed, permeabilized and labeled with anti-Lamp1 antibodies, followed byCy3-F(ab')$_2$ anti-rat IgG antibodies. The yellow arrowhead indicates a 7 μm-wire surrounded by Lamp1 staining.*

To analyze the subcellular localization of the wires in NIH/3T3, cells were fixed and stained with antibodies to detect the Lysosomal Associated Membrane Protein (Lamp1), a marker of late endosomal/lysosomal endosomes. Fig. 1c displays a close-up immunofluorescent image depicting a 7 μm-wire surrounded by a Lamp1 tagged membrane. A 3D reconstruction of the data is illustrated in Supporting Information (S6). The proportion of LAMP1-positive elongated compartments with respect to the total number of internalized wires was estimated at $14 \pm 5$ %, an outcome that was in line with the TEM data. These results suggest that a large fraction of internalized wires are dispersed in the cytosol 24 h after internalization, and that they are able to probe the dynamical properties of the intracellular medium. To complete these studies, toxicity assays were carried out at levels up to 60 wires per cell to evaluate the mitochondrial activity, the cell proliferation and the production of reactive oxygen species (S7). These surveys show that even at the highest content, the wires display no acute short-term (< 100 h) toxicity [42-45].

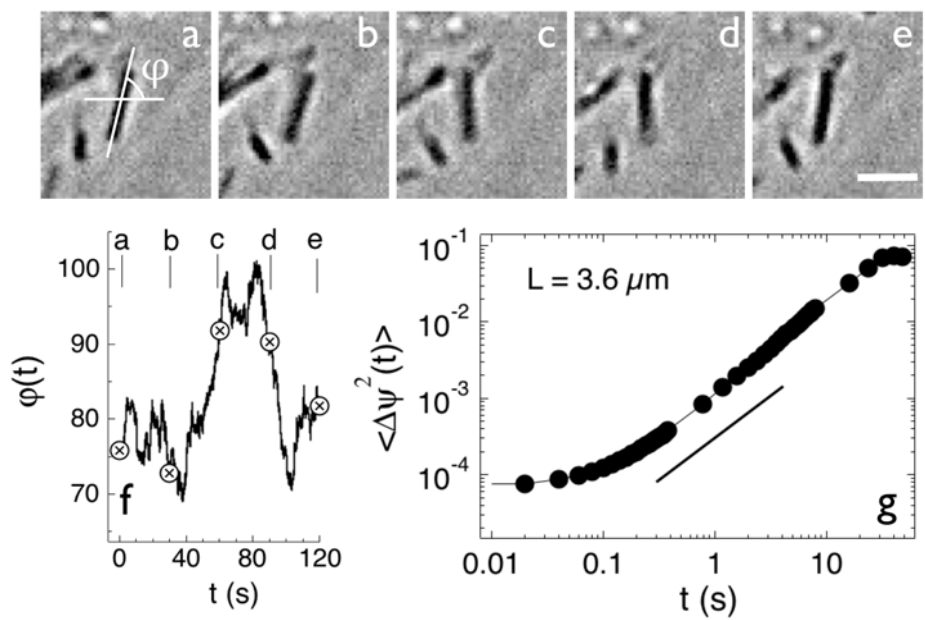

**Figure 2:** *a-e) Images of 3.6 µm long wire internalized in a NIH/3T3 fibroblast at time intervals 0, 30, 60, 90 and 120 s respectively. In a), the angle $\varphi(t)$ determines the orientation of the wire, and the scale bar is 3 µm. f) Time dependence of the angle $\varphi(t)$ corresponding to the stack of images (a-e). Mean-squared angle displacement (MSAD) of the 3.6 µm long wire as a function of the time. The $\langle\Delta\psi^2(t)\rangle$-profile exhibits a plateau at short time followed by an increase at longer times. The straight line in g) corresponds to a linear increase $\langle\Delta\psi^2(t)\rangle\sim t$.*

## 3.2. Mean-Squared Angular Displacement (MSAD)

For passive microrheology, NIH/3T3 fibroblast cells were incubated with magnetic nanowires of length 4 µm and at 10 wires per cell. Because of their dispersity, the fibroblasts were exposed to wires of length comprised between 1 and 10 µm. In all assays, the cells were found to maintain their morphology and adherence properties following a 24 h exposure. The fluctuations of the wires were monitored by optical microscopy using a ×100 objective (numerical aperture 1.3) and recorded using a fast digital camera at the rate of 100 to 1000 images per second. Typical movies were made of 10000 images. Figs. 2a-e display a 3.6 µm long internalized wire at different time intervals, 0, 30, 60, 90 and 120 s respectively. The wire orientation was determined by the angle $\varphi(t)$ defined with respect to the horizontal axis (Fig. 2a). The diagram displaying the spherical coordinate system used for the description of the wire orientation is provided in Supporting Information (S5). Fig. 2f shows the time dependence of the angle $\varphi(t)$ corresponding to this stack of images. During the 120 s of the experiment, the angle fluctuates randomly around an average value of 85°, and shows excursions of ± 15° with respect to the mean. The out-of-plane motion of the wire was extracted from their 2-dimensional projection according a procedure described earlier [33]. The variable $\psi(t)$, which time derivative varies as $\dot{\psi} = \sin\theta\,\dot{\varphi}$ was calculated using the actual and the apparent lengths of the wire, $L$ and $L_{app}(t) = L\sin\theta(t)$ respectively. On a longer time scale, the wire explored different orientational configurations and from this recording it is possible to derive its actual length $L$. In purely viscous fluids such as water-glycerol mixtures, the mean-squared angular displacement associated to the variable $\psi(t)$ was found to vary linearly with time and to correctly predict the orientational diffusion coefficient and the viscosity of the fluid [33]. Detailed calculations of the MSAD noted $\langle\Delta\psi^2(t)\rangle$ are also given in Supporting Information (S5 and S8). Fig. 2g shows the MSADs of the 3.6 μm long wire internalized into a fibroblast as a function of the time. The $\langle \Delta \psi^2(t) \rangle$-profile exhibits a plateau at short time ($t < 0.1$ s) followed by an increase at longer times ($t > 0.5$ s). In the following, these two time regimes will be identified as Regime I and Regime II, respectively.

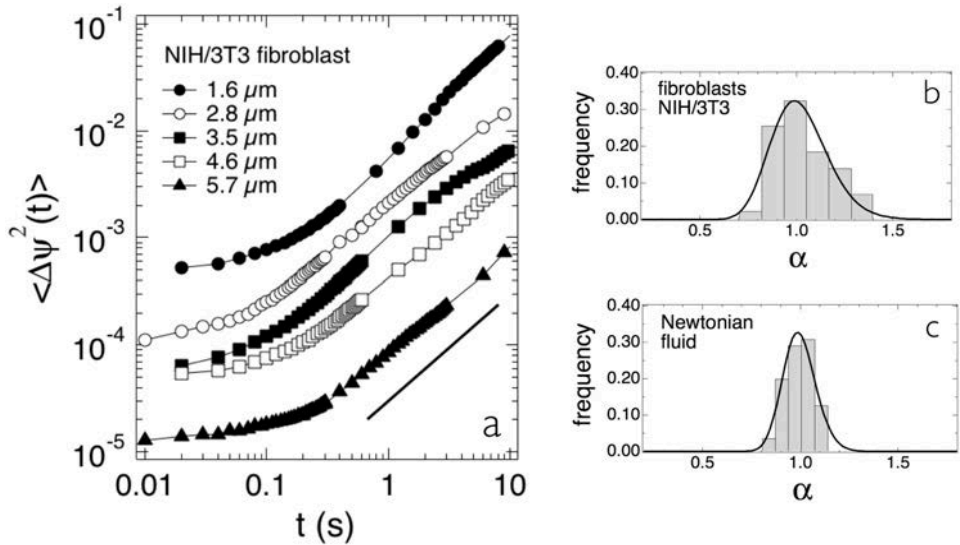

***Figure 3: a)*** *Mean-squared angular displacement (MSAD) of wires of length $L = 1.6$ to 5.7 μm embedded in the intracellular medium of NIH/3T3 fibroblasts ($T = 37$ °C). At times between 0.5 s and 10 s, the MSAD was adjusted by a power law of the form $\langle \Delta \psi^2(t) \rangle \sim t^{\alpha}$. The distribution of exponents $\alpha$ is shown in **b)** for the cells and in **c)** for a water-glycerol mixture. The plateau observed in the MSADs for short times reflects the limit in angular resolution.*

### 3.3. Mean-Squared Angular Displacements as a function of the wire length

The previous measurement was repeated with wires of different lengths. Fig. 3a displays the MSADs of 5 wires internalized into fibroblasts as a function of the time ($T = 37$ °C), their lengths varying between 1.6 and 5.7 μm. The features identified previously as well as Regimes I and II were again observed. Interestingly, the MSAD were found to depend strongly on the length: the shorter the wire, the larger their rotational fluctuations. In the following, Regimes I and II are discussed separately.

In Regime I, the MSADs extrapolated at $t \to 0$ were estimated as a function of the wire length. Displayed in Fig. 4, the $\langle \Delta \psi^2(t \to 0) \rangle$-data for the NIH/3T3 cells exhibit a strong decrease with increasing $L$. This dependence is compared to the minimum MSAD detectable by the present technique, and noted $2 \varepsilon_{Rot}^2(L)$ [24, 46]. The angular resolution was determined from separate experiments in which the rotational fluctuations of wires dispersed in a 92 vol. % water/glycerol solution ($\eta = 0.45$ Pa s) were studied. The data on the glycerol-water mixture are shown as a continuous line in Fig. 4 and corresponds to the expression:

$$2 \varepsilon_{Rot}^2(L) = 7.5 \times 10^{-24} L^{-3.5} \tag{1}$$

$L$ being in meter and $\varepsilon_{Rot}$ in radian. The expression in Eq. 1 also provides the angular resolution of the wire-based micro-rheology technique: it gives $\varepsilon_{rot} = 1.3°$ for a 2 μm wire and $\varepsilon_{rot} = 0.02°$ for a 20 μm-wire. In Fig. 4, the good agreement between the $\langle \Delta \psi^2(t \to 0) \rangle$ data collected on the cells and $2\varepsilon_{Rot}^2$ indicates that the upturn toward the plateau regime in Regime I is related to the resolution of the present technique, and not to another phenomenon, such as the elasticity of the intracellular network [47].

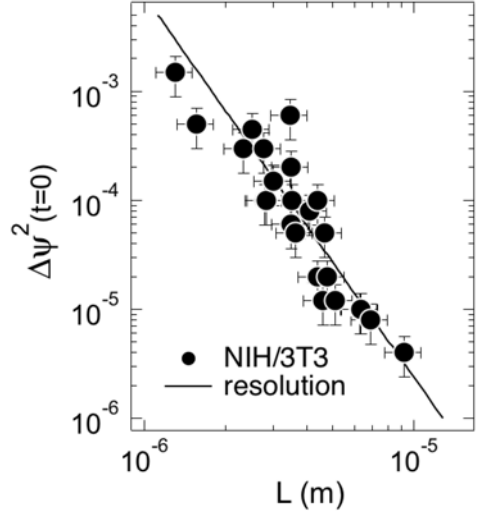

**Figure 4**: *Comparison between the MSADs extrapolated to zero in NIH/3T3 fibroblasts and the minimum MSAD detectable by the present micro-rheology technique. The straight line corresponds to the power law* $2\varepsilon_{Rot}^2(L) = 7.5\times10^{-24}L^{-3.5}$.

At longer times (Regime II), *i.e.* between 0.5 and 10 s, an increase of the MSADs *versus* time was observed. To ascertain the characteristics of this increase, the MSADs were adjusted using a power law of the form $\langle \Delta\psi^2(t) \rangle \sim t^\alpha$, $\alpha$ being an adjustable parameter. The distribution of exponents for the fibroblasts is shown in Fig. 3b, and compared to those found for a Newton fluid (water/glycerol mixture) in Fig. 3c. For the water/glycerol mixture, the center $\alpha_0$ of the distribution was 0.99 and the polydispersity $s_\alpha$ 0.084. $\alpha_0$-values close to 1 are indeed expected in the Newtonian regime. For the NIH/3T3 fibroblasts, the exponent distribution was again peaked around 1 ($\alpha_0 = 1.00$) and the polydispersity was slightly higher ($s_\alpha = 0.148$). These results evidence the existence of a diffusive regime in the rotational fluctuations of wires embedded inside NIH/3T3 fibroblasts.

### 3.4. Rotational diffusion constant *versus* length

Based on the previous analysis, the mean-squared angular displacement (Regime II) was adjusted using [33]:

$$\langle \Delta\psi^2(t, L) \rangle = 2D_{rot}(L)t \tag{2}$$

where $D_{rot}(L)$ is the rotational diffusion coefficient of the wire. For a cylinder of length $L$ and diameter $d$ embedded in a viscous fluid of viscosity $\eta$, $D_{rot}(L)$ is given by:

$$D_{rot}(L) = \frac{3k_B T}{\pi L^3 \eta} g\left(\frac{L}{d}\right) \tag{3}$$

where $k_B T$ the thermal energy, $k_B$ Boltzmann constant and $g(L/d)$ is a dimensionless function of the anisotropy ratio $p = L/d$. In this study, we assume that $g(p) = \ln(p) - 0.662 + 0.917\,p - 0.050\,p^2$ which is valid in the interval $2 < p < 20$ [48]. From the values of rotational diffusion coefficients, it is in principle possible to deduce a viscosity $\eta$ (Eq. 3). However, Eqs. 2 and 3 derive from the fluctuation-dissipation relation that is valid only for equilibrium systems [6, 7, 49]. It is not expected to hold for the interior of cells, where thermal and non-thermal active forces are both playing a role in diffusion processes [16, 17]. In the following, we examine the $L$-dependence of the rotational diffusive constants derived from the MSADs. This analysis implies that the intracellular medium is regarded as a medium of effective viscosity η, which includes both active and non-active contributions to the intracellular dynamics. The rotational diffusion coefficient was estimated from Regime II and plotted against the wire length in Fig. 5a. It was found to decrease with increasing $L$ according to $D_{Rot}(L) \sim g\left(\frac{L}{d}\right)/L^3$ using $d = 400$ nm (Eq. 3, continuous lines). The dispersion of the data with respect to the calculated curve in Fig. 5a is more important for the fibroblasts than for the model fluids. For the glycerol/water mixtures, this scattering was attributed to the diameter distribution of the wires [33]. In Figs. 5, the data suggest the existence of an additional broadening mechanism arising from the heterogeneities of the intracellular medium or from the cell-to-cell variability [4, 5]. The adjustment with the theoretical expression results in an effective viscosity of $\eta = 0.16$ Pa s, $i.e.$ 200 times the viscosity of water at this temperature. The $1/L^3$-scaling observed for the rotational diffusion coefficient indicates moreover that the intracellular viscosity does not depend on the length scale in the range $L = 1 - 10\ \mu m$. Note that in our experiments, correlations between the values of the effective viscosity, the locations of the wires in the cytoplasm ($i.e.$ close to the plasma membrane or to the nucleus) and the length of the wires were not observed.

To ensure that the data in Fig. 5a were not specific to the NIH/3T3 fibroblasts, additional experiments were carried out on MDCK canine kidney epithelial cells. Measurements of MSADs $versus$ time were found to exhibit the same Regime I and II as those of Fig. 3a, $i.e.$ a plateau at short times followed by a $\langle \Delta \psi^2(t) \rangle \sim t^\alpha$ behavior, where $\alpha$ was again distributed and close to 1. As for the fibroblasts, the rotational diffusion constant scaled as $D_{rot}(L) \sim g\left(\frac{L}{d}\right)/L^3$, yielding here an effective viscosity $\eta_0 = 1.40$ Pa s, $i.e.$ 1800 times the viscosity of water at body temperature. From these supplementary experiments, it can be concluded that the wire-based micro-rheology technique provides a direct method to measure the effective viscosity of the intracellular medium.

### 3.5. Spherical beads $versus$ wires

For sake of completeness, passive micro-rheology using beads as intracellular probes was performed. The time dependence of the mean-squared displacements was derived and compared with that of the MSADs of wires. Spherical poly(styrene) beads of 1 μm in diameter (Molecular Probes, Invitrogen) were incubated with the NIH/3T3 fibroblasts and their mean-squared displacements monitored by microscopy in the same conditions as for the wires. With the beads, the MSDs exhibited power laws and exponents comprised between 1 and 1.6 in the time range 10 ms – 10 s (S9). The results with the beads were also in excellent agreement with those of the literature [4, 5, 13, 14, 16, 19].

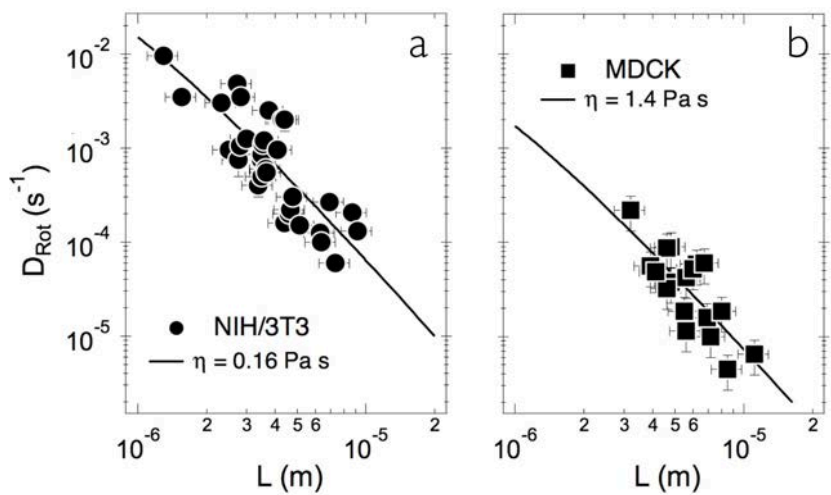

***Figure 5:*** *Rotational diffusion constant $D_{Rot}(L)$ as a function of the length for wires dispersed in* ***a)*** *NIH/3T3 fibroblasts and* ***b)*** *MDCK canine kidney epithelial cells ($T = 37\,°C$). The continuous line results from least-square calculations using Eq. 3.*

With the fibroblasts, no diffusive dynamics could be evidenced, and a value of an effective viscosity could not be obtained. Another interesting feature of the bead micro-rheology was the broad bead-to-bead variability in the time responses, also described in the literature and illustrated in Supplementary Information (S9) [12-16]. The specificity of the wires as compared to other micro-rheology probes might indicate a different type of coupling of the anisotropic objects with the cytoskeleton. This comparison demonstrates the benefits of the wire-based microrheology, and also emphasizes that the two techniques lead to different results. The differences between the intracellular behaviors of beads and wires can have several origins. One plausible explanation could be that due to their large aspect ratio, wires probe the surrounding environment over larger length scales, averaging the structural heterogeneities of the medium. Another explanation relates to the fact that contrary to the beads, wires are mostly dispersed in the cytosol [36].

## 4. Conclusion

In this study, we evaluate the potential applications of wires as alternative probes for investigating the mechanical properties of the intracellular medium. Wires of 400 nm in diameter and 1 to 10 μm in length were synthesized by electrostatic complexation. The co-assembly process followed a brick-and-mortar strategy where the elementary bricks are sub-10 nm iron oxide nanoparticles and the mortar comprised of charged polymers. Passive micro-rheology was implemented on living cells to study the rheological response of the intracellular medium. The advantages of using wires are that they are easily internalized by live cells, non-toxic and dispersed into the cytosol. The dynamics of the internalized wires was studied in terms of mean-squared angular displacement. Two time regimes were identified. The MSAD exhibits a plateau at short time ($t < 0.1$ s), which is followed by an increase at longer times ($t > 0.5$ s). The first regime was ascribed to an effect of angular resolution of the image processing tracking technique. In this second regime, the MSAD was associated with a diffusive-like behavior, where it increases linearly with time. The rotational diffusion constant was found to scale with $1/L^3$ ($L$ being the length of the wire) on more than two orders of magnitude and for two distinctive cell lines (fibroblasts and epithelial cells). The present data are interpreted in terms of an effective intracellular viscosity that include thermal and non-thermal active forces. The values of the viscosities found for the fibroblasts and for the epithelial cells were 160 and 1800 times that of water. The results also suggest that the intracellular viscosity does not depend on the length scale in the investigated range in the range $L = 1 - 10 \ \mu m$. In conclusion, the present work demonstrates that micron-sized wires are well suited to study the mechanics of complex and/or heterogeneous fluids of small volumes (< 1 pL). Their field-induced magnetization properties make them unique objects for magnetic actuation in microfluidics and nonlinear micro-rheology.

## Acknowledgments

We thank Armelle Baeza-Squiban, Patricia Bassereau, Jean-Paul Chapel, Jérôme Fresnais, Jean-Pierre Henry, Eléna Ishow, Florence Niedergang, Alain Richert, Malak Safi, Olivier Sandre, Nicolas Schonbeck, Daphne Weihs, Minhao Yan for fruitful discussions. The Laboratoire Physico-chimie des Electrolytes, Colloïdes et Sciences Analytiques (UMR Université Pierre et Marie Curie-CNRS n° 7612) is recognized for providing us with samples and characterization of the magnetic nanoparticles. We extend our thanks to the team Phagocytosis and Bacterial Invasion from Institut Cochin (Paris) for the immunofluorescence and localization assays.

## Supporting Information

In the Supporting Information, the characterization of the iron oxide nanoparticles and that of the magnetic nanowires are provided in S1-S3 to S4, respectively. S5 describes the spherical coordinate system for diffusing wires and the theoretical approach for the calculation of the mean-squared angular displacement. S6 illustrates the 3D reconstruction of the immunofluorescence data for internalized wires, whereas MTT toxicity assays for the particles and for the wires are described in S7. The assessment of the present wire-based micro-rheology technique in terms of resolution is discussed in S8. S9 provides data on the 1 μm beads internalized in NIH/3T3 fibroblasts. Finally, movie#1 illustrates the fluctuations of a wire as a function of the time. This information is available via the Internet at on the website of the journal.

## References

1.	Bao G, Suresh S. Cell and molecular mechanics of biological materials. Nat. Mater. 2003;2:715-725.
2.	Kuimova MK, Botchway SW, Parker AW, Balaz M, Collins HA, Anderson HL, et al. Imaging intracellular viscosity of a single cell during photoinduced cell death. Nature Chemistry 2009;1:69-73.
3.	Ellis RJ. Macromolecular crowding: Obvious but underappreciated. Trends Biochem.Sci. 2001;26:597-604.
4.	Weihs D, Mason TG, Teitell MA. Bio-microrheology: A frontier in microrheology. Biophys. J. 2006;91:4296-4305.
5.	Panorchan P, Lee JSH, Daniels BR, Kole TP, Tseng Y, Wirtz D. Probing cellular mechanical responses to stimuli using ballistic intracellular nanorheology. In: Wang YL, Discher DE, editors. Cell mechanics. San Diego: Elsevier Academic Press Inc, 2007. p. 115-140.
6.	MacKintosh FC, Schmidt CF. Active cellular materials. Curr. Opin. Cell Biol. 2010;22:29-35.
7.	Squires TM, Mason TG. Fluid mechanics of microrheology. Annu. Rev. Fluid Mech. 2010;42:413-438.

8.      Park S, Koch D, Cardenas R, Kas J, Shih CK. Cell motility and local viscoelasticity of fibroblasts. Biophys. J. 2005;89:4330-4342.

9.      Brunner C, Niendorf A, Kas JA. Passive and active single-cell biomechanics: A new perspective in cancer diagnosis. Soft Matter 2009;5:2171-2178.

10.     Mitrossilis D, Fouchard J, Guiroy A, Desprat N, Rodriguez N, Fabry B, et al. Single-cell response to stiffness exhibits muscle-like behavior. Proceedings of the National Academy of Sciences of the United States of America 2009;106:18243-18248.

11.     Mitrossilis D, Fouchard J, Pereira D, Postic F, Richert A, Saint-Jean M, et al. Real-time single-cell response to stiffness. Proceedings of the National Academy of Sciences of the United States of America 2010;107:16518-16523.

12.     Fabry B, Maksym GN, Butler JP, Glogauer M, Navajas D, Fredberg JJ. Scaling the microrheology of living cells. Physical Review Letters 2001;87:148102.

13.     Tseng Y, Kole TP, Wirtz D. Micromechanical mapping of live cells by multiple-particle-tracking microrheology. Biophys. J. 2002;83:3162-3176.

14.     Lee JSH, Panorchan P, Hale CM, Khatau SB, Kole TP, Tseng Y, et al. Ballistic intracellular nanorheology reveals rock-hard cytoplasmic stiffening response to fluid flow. J. Cell Sci. 2006;119:1760-1768.

15.     Hoffman BD, Massiera G, Van Citters KM, Crocker JC. The consensus mechanics of cultured mammalian cells. Proceedings of the National Academy of Sciences of the United States of America 2006;103:10259-10264.

16.     Gallet F, Arcizet D, Bohec P, Richert A. Power spectrum of out-of-equilibrium forces in living cells: Amplitude and frequency dependence. Soft Matter 2009;5:2947-2953.

17.     Robert D, Nguyen TH, Gallet F, Wilhelm C. In vivo determination of fluctuating forces during endosome trafficking using a combination of active and passive microrheology. Plos One 2010;5:e10046.

18.     Fakhri N, MacKintosh FC, Lounis B, Cognet L, Pasquali M. Brownian motion of stiff filaments in a crowded environment. Science 2010;330:1804-1807.

19.     Yizraeli ML, Weihs D. Time-dependent micromechanical responses of breast cancer cells and adjacent fibroblasts to electric treatment. Cell Biochemistry and Biophysics 2011;61:605-618.

20.     Saxton MJ, Jacobson K. Single-particle tracking: Applications to membrane dynamics. Annu. Rev. Biophys. Biomolec. Struct. 1997;26:373-399.

21.     Balland M, Desprat N, Icard D, Fereol S, Asnacios A, Browaeys J, et al. Power laws in microrheology experiments on living cells: Comparative analysis and modeling. Physical Review E 2006;74:17.

22.     Deng LH, Trepat X, Butler JP, Millet E, Morgan KG, Weitz DA, et al. Fast and slow dynamics of the cytoskeleton. Nature Materials 2006;5:636-640.

23.     Cheng Z, Mason TG. Rotational diffusion microrheology. Phys. Rev. Lett. 2003;90:018304.

24.     Cheong FC, Grier DG. Rotational and translational diffusion of copper oxide nanorods measured with holographic video microscopy. Opt. Express 2010;18:6555-6562.

25.     Dhar P, Cao YY, Fischer TM, Zasadzinski JA. Active interfacial shear microrheology of aging protein films. Phys. Rev. Lett. 2010;104.

26.     Celedon A, Hale CM, Wirtz D. Magnetic manipulation of nanorods in the nucleus of living cells. Biophys. J. 2011;101:1880-1886.

27.     Frka-Petesic B, Erglis K, Berret J-F, Cebers A, Dupuis V, Fresnais J, et al. Dynamics of paramagnetic nanostructured rods under rotating field. Journal of Magnetism and Magnetic Materials 2011;323:1309-1313.

28.     Xiao L, Qiao Y, He Y, Yeung ES. Imaging translational and rotational diffusion of single anisotropic nanoparticles with planar illumination microscopy. J. Am. Chem. Soc. 2011;133:10638–10645.

29.     Cappallo N, Lapointe C, Reich DH, Leheny RL. Nonlinear microrheology of wormlike micelle solutions using ferromagnetic nanowire probes. Physical Review E 2007;76:6.

30.     Fung AO, Kapadia V, Pierstorff E, Ho D, Chen Y. Induction of cell death by magnetic actuation of nickel nanowires internalized by fibroblasts. The Journal of Physical Chemistry C 2008;112:15085-15088.

31.     Chippada U, Yurke B, Georges PC, Langrana NA. A nonintrusive method of measuring the local mechanical properties of soft hydrogels using magnetic microneedles. Journal of Biomechanical Engineering-Transactions of the Asme 2009;131:021014.

32.     Han Y, Alsayed AM, Nobili M, Zhang J, Lubensky TC, Yodh AG. Brownian motion of an ellipsoid. Science 2006;314:626-630.

33.     Colin R, Yan M, Chevry L, Berret J-F, Abou B. 3d rotational diffusion of micrometric wires using 2d video microscopy. Europhysics Letters 2012;97:30008.

34.     Fresnais J, Berret J-F, Frka-Petesic B, Sandre O, Perzynski R. Electrostatic co-assembly of iron oxide nanoparticles and polymers: Towards the generation of highly persistent superparamagnetic nanorods. Advanced Materials 2008;20:3877-3881.

35.     Yan M, Fresnais J, Sekar S, Chapel JP, Berret J-F. Magnetic nanowires generated via the waterborne desalting transition pathway. Acs Applied Materials & Interfaces 2011;3:1049-1054.

36.     Safi M, Yan MH, Guedeau-Boudeville MA, Conjeaud H, Garnier-Thibaud V, Boggetto N, et al. Interactions between magnetic nanowires and living cells: Uptake, toxicity, and degradation. Acs Nano 2011;5:5354-5364.

37.     Massart R, Dubois E, Cabuil V, Hasmonay E. Preparation and properties of monodisperse magnetic fluids. J. Magn. Magn. Mat. 1995;149:1 - 5.

38.     Berret J-F, Sandre O, Mauger A. Size distribution of superparamagnetic particles determined by magnetic sedimentation. Langmuir 2007;23:2993-2999.

39.     Berret J-F, Sehgal A, Morvan M, Sandre O, Vacher A, Airiau M. Stable oxide nanoparticle clusters obtained by complexation. Journal of Colloid and Interface Science 2006;303:315-318.

40.     Berret J-F. Stoichiometry of electrostatic complexes determined by light scattering. Macromolecules 2007;40:4260-4266.

41.     Hultgren A, Tanase M, Felton EJ, Bhadriraju K, Salem AK, Chen CS, et al. Optimization of yield in magnetic cell separations using nickel nanowires of different lengths. Biotechnol. Prog. 2005;21:509 - 515.

42.     Doshi N, Mitragotri S. Needle-shaped polymeric particles induce transient disruption of cell membranes. Journal of the Royal Society Interface 2010;7:S403-S410.

43.     Gratton SEA, Ropp PA, Pohlhaus PD, Luft JC, Madden VJ, Napier ME, et al. The effect of particle design on cellular internalization pathways. Proceedings of the National Academy of Sciences of the United States of America 2008;105:11613-11618.

44.     Nan AJ, Bai X, Son SJ, Lee SB, Ghandehari H. Cellular uptake and cytotoxicity of silica nanotubes. Nano Lett. 2008;8:2150-2154.

45.     Song MM, Song WJ, Bi H, Wang J, Wu WL, Sun J, et al. Cytotoxicity and cellular uptake of iron nanowires. Biomaterials 2010;31:1509-1517.

46.     Savin T, Doyle PS. Static and dynamic errors in particle tracking microrheology. Biophys. J. 2005;88:623-638.

47.     van Zanten JH, Rufener KP. Brownian motion in a single relaxation time maxwell fluid. Physical Review E 2000;62:5389-5396.

48.     Tirado MM, Martinez CL, Delatorre JG. Comparison of theories for the translational and rotational diffusion-coefficients of rod-like macromolecules - application to short DNA fragments. J. Chem. Phys. 1984;81:2047-2052.

49.     Abou B, Gallet F, Monceau P, Pottier N. Generalized einstein relation in an aging colloidal glass. Physica a-Statistical Mechanics and Its Applications 2008;387:3410-3422.